\documentstyle[12pt]{article}
\begin{document}                                                             
\title{What heavy quanta bounds could be inferred from a
Higgs discovery? }  
\author{ J.~Lorenzo~Diaz-Cruz \\
Instituto de Fisica, BUAP, 72570 Puebla, Pue, Mexico }
\maketitle
\begin{abstract}
The Higgs-gauge bosons couplings
$g_{hVV}$  ($V=g,\gamma,W,Z$) receive non-decoupling 
corrections due to heavy quanta, and deviations 
from the SM predictions can be used to test its presence.
The possible Higgs signal recently reported at LEP,
with $m_h=115$ GeV, severely constrains the presence  
of heavy quanta, such as a heavy fourth family. At Tevatron,
the Higgs production by gluon fusion, followed by the decay 
$h \to WW^*$, 
can also be used to probe the existence of heavy 
colored particles, including additional  standard or mirror families, 
chiral colored sextets and octet quarks. 
Within the MSSM, we also find that gluon fusion is a sensitive
probe for the spectrum of squark masses. 

\end{abstract}


\bigskip

\newpage

{\bf {1.- Introduction.}}
The discovery of the Higgs boson as the remmant of the mechanism of
electro-weak symmetry breaking (EWSB), is one of the most cherised
goals of present and future high-energy experiments.
Within the minimal standard model (SM), the mass of the
physical Higgs particle is a free-parameter, but present indirect 
data seems 
to favor a moderate mass ($105 < m_h < 220$ GeV) \cite{higgsrev},  
however this conclusion may be changed by the presence
of  new phyics with an scale $\Lambda \simeq O(1) $ TeV
\cite{Meandthem}.  On the other hand, the minimal supersymmetric 
version of the SM (MSSM), which is one of its most appealing extensions,
predicts a light Higgs boson, with
an upper mass bound of about  130 GeV \cite{carenaetal}.
Detection of the full spectrum of Higgs bosons in the SM and beyond,
constitute an important test of the possible realization of the Higgs 
mechanism in fundamental physics. In fact, a few events that could be 
interpreted as a SM Higgs signal were reported at the last stages 
of LEP \cite{newsofhiggs}.

 The characteristic Higgs boson couplings determine the
strategies employed for its search
at present and future colliders. For instance, at LEP 
the reported Higgs events would come from the reaction
$e^+e^- \to Z+h$, which takes advantage of
the large couplings of the Higgs with massive gauge bosons. 
At Hadron colliders one can also test
these couplings, either through the reactions 
$pp \to W+h, Z+h$, or through the decays $h\to WW,ZZ$;
vector fusion (WW and ZZ) can also be used for heavy Higgs masses.
The couplings of the Higgs with the heavier fermions, can be studied
either by open production ($t\bar{t}h, b\bar{b}h$) \cite{ourhix}
or by the loop-induced Higgs coupling
with gluon and photon pairs ($hgg, h\gamma\gamma$) \cite{hixhunter}. 
Any additional heavy particle that receives its
mass from the SM Higgs mechanism, will
couple to the Higgs with strength proportional to the particle 
mass itself, and
will induce non-decoupling contributions to the
1-loop vertices $hgg,\,  h\gamma\gamma$.
These effects can play a significant role and may
be probed at future colliders,  as it has been
explored in previous studies \cite{chanofourth}. However,
 the presence of these new particles will
also induce non-decoupling corrections to the tree vertices
$hf\bar{f}$, $hWW$ and $hZZ$ \cite{veltmanetal}, which can affect the 
decay rate of detectable signatures, and thus  must be included in the 
analysis.
Beyond the SM, the lepton flavor violating Higgs decays
$h\to l_i l_j$ could also be tested at colliders
\cite{myhlfv}.

In this letter we are interested in studying the possible bounds 
on the presence of additional colored particles, that can be infered
from the possible discovery of a Higgs signal, which then
can become another tool for precision electro-weak studies.
We consider first the possibility that
the reported LEP Higgs signal is indeed true, and use it
to constrain the presence of heavy quanta.
Then, at hadron colliders we study the Higgs signal coming from 
the production
by gluon fusion followed by the decay $h\to WW^*$, and find
the bounds on heavy quanta that can be obtained from Tevatron (RUN-II). 
We also study gluon fusion for the MSSM higgs
bosons, as a possible test for the mass spectrum
of squarks; large effects are obtained when the loop amplitude includes
non-degenerate squark masses, which could be realized in 
many particular models \cite{polonskypom}.

{\bf {2.- Heavy quanta effects on Higgs vertices.}}
In order to probe heavy scales through their effect on
the Higgs coupling, we shall consider extensions 
of the SM that include additional particles, but with a 
minimal Higgs sector consisting of one doublet.
The possible representations can only include fermion doublets and singlets 
\cite{chiralferms}, which can be arranged into:
a) additional families with SM quantum numbers
b) additional families with mirror properties,
c) a combination of the above.
Within cases a and b, new quarks must have the same color properties
as in the SM, but within case c they could lay in larger $SU(3)_c$
representations, like sextets or octets.

Since these new states have electroweak charges, they will contribute
in general to the parameters S,T,U, but not necessarily. 
Present global fits for electroweak data seem to exclude
more than one additional SM family \cite{erlerlangack}, 
however this conclusion relies
on the assumption that no other physics occurs at the mass scale
of the new fermions, which may not be the case in specific models,
for instance in SUSY models. For illustration, we can
consider an extended SUSY model with an additional (4th) family, 
and assume a mass degeneracy for both components of the 
fermion doublets, this gives $T=0$;
whereas fermions give $S=2 N_c/6\pi$, the contribution of sfermions
to $S$ is given by 
$S=\frac{N_c}{36 \pi}log(\frac{m^2_{\tilde u_{Li}} }{m^2_{\tilde d_{Li}} })$
\cite{siannah}, 
thus by choosing the appropriate masses one could decrease the total
value of S, and satisfy present experimental constraints.  
Furthermore, including neutral majorana particles
gives a negative contribution to S, and it makes even easier to satisfy
present bounds. The light Higgs boson of the model $h^0$
behaves like the SM one in the limit $m_A >> m_Z$, under which the 
heavy Higgs bosons decouple.

Thus, we can identify two scenarios for our analysis, 
in Scenario-I we shall  include a specific case of heavy quanta,
namely a heavy fourth family,
for which we can evaluate all the corrections induced   
on the vertices $hq\bar{q}$, $hgg/h\gamma\gamma$ and $hWW/hZZ$,
using the formalism of radiative corrections.
 The second case (Scenario-II) includes the situation when the mass of 
heavy colored particles comes  from the SM Higgs boson, but which are weakly 
bounded by electroweak precision meassurements, i.e. we simply assume that 
there is some unspecified physics at the mass scale of the
new particles that makes their contribution to the Peskin-Takeuchi
parameters $S,T$ to be within experimental range \cite{peskintak}.

For both scenarios, where we have additional heavy particles that receive 
their mass from the SM Higgs mechanism, but are weakly constrained by 
electroweak precision data, the study of the Higgs signal can test the 
presence of such heavy particles.
 To describe the effects of heavy quanta on the higgs couplings
to gauge bosons, we shall write them in terms of the SM values
($ g^{SM}_{hVV} $) as: 
$g_{hVV}= g^{SM}_{hVV} (1+\epsilon_V)$,
where $V=W,Z,g,\gamma$, and similarly for the fermion couplings
$g_{hff}= g^{SM}_{hff} (1+\epsilon_f)$. The parameters $\epsilon_{f,V}$, 
encompass the effects of heavy quanta.

We could also consider effects arising from heavy particles  
that receive their mass from the breaking of new symmetries 
characterized by a large scale $V_{new} >> v=246$ GeV, induced by another 
set of heavy Higgs bosons ($\Phi_{new}$). Supposse this
is communicated to the SM Higgs by including in the scalar potential
a mixing term  of the form:  $\lambda |\Phi_{new}|^2 |\phi_{sm}|^2$.
Then one can also include the contribution of these ultra-heavy
particles to the parameters $\epsilon_V$.
However, one can easily verify that in this case the mixing angle 
that relates the weak and mass-eigenstate  Higgs basis,
is suppressed by the large scale ($V_{new}$), and it induces
decoupling effects at low-energies, namely 
$\epsilon_g \simeq v/V_{new}$. 

{\bf {3.-Bounds on heavy quanta from the LEP Higgs signal.}}
The possible discovery of a Higgs signal, with $m_h=115$ geV,
could be a remarkable closing for the series of sucessfull 
experiments performed at LEP. The reported events are indeed consistent
with a SM higgs interpretation \cite{newsofhiggs}, though a definite 
conclusion will have
to wait for the coming stages of Tevatron RUN-II or even the LHC.
Here, we shall assume that the Higgs signal is indeed real, and will
obtain bounds on heavy quanta for Scenario-I.  
For the simple case of a heavy fourth family, 
one finds an interesting relation between $\epsilon_{f,W,Z}$
and the corresponding expression for the  parameter 
$T$, namely, 
\begin{eqnarray}
\epsilon_b&=& \frac{\alpha T}{2}+ \frac{N_c G_F }{12\sqrt{2} \pi^2 } 
                    (m^2_1+m^2_2) \ \nonumber \\
\epsilon_Z&=& -\frac{N_c G_F }{6 \sqrt{2} \pi^2 }  (m^2_1+m^2_2) \nonumber \\
\epsilon_W&=& \epsilon_Z + \frac{\alpha T }{2}
\end{eqnarray}
where $T$ includes the contribution from the fourth family fermions, with 
masses $m_{1,2}$.
It is interesting to comment that this relation holds
for any other colored heavy family that transform as
doublets and singlets under $SU(2)_L\times U(1)_Y$.

Then, to evaluate the event rate for Higgs production at LEP,
we find convenient to write the product of the
cross-section for $e^+e^- \to ZH$ times the b.r. of $h\to bb$, in terms 
of the SM result, namely: 
$[\sigma \times B.R.(h\to bb)]_{new}= 
R_{Zb} \times [\sigma \times B.R.(h\to bb)]_{SM}$,
where $R_{Zb}$ is given by 
\begin{equation}
R_{Zb}  =   \frac{(1+ \epsilon_Z)^2(1+\epsilon_b)^2 }
       {[ 1+\sum_Y (\epsilon^2_Y + 2 \epsilon_Y)*B.R.(h_{SM} \to YY)]},
\end{equation}  
the sum in the denominator runs over the allowed
modes ($Y=t,b,g,A,W,Z$).
Results for $R_{Zb}$ corresponding to a heavy fourth quark 
generation, and  $m_h=115$ GeV, are shown in Table 1. Then,  
by comparing the event rate reported at LEP, which seems consistent
with the SM Higgs (for which $R_{Zb}=1$), and the value for $R_{Zb}$
predicted in the presence of the heavy quanta, one could test their presence.
From Table 1, we can see that by just having a 25\% 
precision on the cross-section determined by
LEP, it would be possible to exclude a heavy fourth generation, 
with masses above 150 GeV, which is a remarkable result.

{\bf {4.- Bounds on heavy quanta at Tevatron.}}
At hadron colliders we are interested in evaluating the effects of
heavy quanta on the Higgs production by gluon fusion. 
In general, the expression for the correction to the
couplings $hgg$ ($\epsilon_{g}$)
is given by a ratio of complicated loop
expresions, however since we are interested in the limit
$m_h << M_{heavy}$, one can use the low-energy theorems \cite{LowETH},
to relate $\epsilon_{g}$ to the beta-function coefficients for 
the strong coupling constant ($\beta_{3,X}$). Then, we have 
$\epsilon_g= \beta_{3,X}/\beta_{3,t}$, where 
$\beta_{3,t (X)}$ denote the contribution of top
and heavy quanta (X) to the strong beta function.
The values of $\epsilon_g$ arising from a pair of heavy color triplets, 
sextets and octet quarks  are : $\epsilon_g= 2,\,  5$ and 6, respectively,
whereas a new SM family plus its mirror partner gives $\epsilon_g= 4$. 
Thus, the new particles
will modify the cross section by gluon fusion,
which can be written in terms of the SM result ($\sigma_{SM}$ )
as:
$\sigma (pp\to h+X) = \sigma_{SM} (1+ \epsilon_g)^2$.
New colored particles give $\epsilon_g > 0$ and
will enhance the corresponding cross-section.

Bounds on the effects of heavy quanta can be obtained at Tevatron 
using gluon fusion and the decay $h\to WW^*$, which was studied in  
detail for the SM in ref. \cite{Hanetal}, this work concluded that it is 
possible to
detect a SM Higgs boson with an integrated luminosity of 30 fb$^{-1}$,
provided that an optimized selection of cuts is implemented; here we shall
only use their first stage of cuts, namely: 
transverse lepton momentum $p_t(e,\mu)>10$ GeV, 
psedorapidity $\eta_{e,\mu} < 1.5$,
lepton invariant mass $m_{ll}>10$ GeV,
jet resolution $\Delta R(l-j) >0.4$,
missing transverse energy $E_T > 10$ GeV. 

For Scenario-I (the fourth family case), the corrections to the 
vertex $hWW$ are negative
(as compared with the positive tree-level value)
and grow with the mass of the heavy quanta; 
for instance for fermion masses of order 500 (700) GeV 
the deviations from the SM tree-level
couplings are of order - 27 (-52) \%, 
which seems to imply that the signal from $h\to WW^*$
will no longer be detectable,
i.e. one can only probe heavy but not ultraheavy quanta.
 With 10 $fb^{-1}$ of integrated luminosity,
it will be possible to exclude only a certain range of
fermion masses, up to about 900 GeV, 
as it is shown in table 1.
On the other hand, the radiative corrections to the
vertex $hb\bar{b}$ show the opposite behaviour, namely they
grow with the heavy fermion masses; however, it will be
difficult to detect this effect at hadron colliders for the
intermediate Higgs masses considered here, since  
$B.R.(h\to b\bar{b})\simeq 1$. Rather than focusing on the
B.R., one has to look for meassuring the Higgs width, which
only seems possible at a muon-collider, where the Higgs would
be produced as an s-channel resonance, which is directly
sensitive to $\Gamma( h\to b\bar{b} )$.

On the other hand, for
the case when the corrections to $\epsilon_{W}$ can be neglected
(Scenario-II), the above cuts already allow us to get interesting bounds 
on the parameter $\epsilon_g$ at  95 \% c.l., 
as it is shown in figure 1 ( for 2 and 10 $fb^{-1}$). 
These bounds will in turn constrain the
presence of heavy colored particles,
provided that the Higgs mass lays in the
intermediate mass range. The interception of the exclusion lines
with the straight dashed line (which corresponds to 
$\epsilon_g=2$), shows the values of Higgs
masses for which it will be possible to limit the presence of a
pair of heavy color triplets. Furthermore, since 
sextets and octet  quarks give a larger value of $\epsilon_g$,
their presence could be excluded too.
It is remarkable that coming stages of Tevatron
will be able to test $\epsilon_g$ at significant levels.

{\bf {5.- Probing a Non-universal sfermion spectrum.}}
The MSSM includes two Higgs doublets,  and the
Higgs spectrum consists of two neutral CP-even scalars $h^0$ and $H^0$,
one CP-odd pseudoscalar $A^0$ and a charged pair $H^\pm$ \cite{himssm}.
The Higgs sector of the model is completely determined at tree level by 
fixing two parameters, conventionally chosen to be $\tan\beta$ 
and the pseudoscalar mass $m_A$. At loop levels, the radiative corrections,
mainly from top and  stop, 
modify the tree-level mass bound ($m_h < m_Z \cos 2\beta$),
allowing to have  $m^{max}_h \simeq 130$ GeV \cite{carenaetal}. 
The corrections to Higgs couplings can also lead to
modification of its production mechanisms 
at hadron colliders, as previously studied 
\cite{adjouadi}. 
Although conventional wisdom states that the contribution of
top/bottom quarks (squarks)  dominate the amplitude for the gluon fusion,
and squarks from first and second generations can be neglected, 
it should be mentioned that this result only holds within the minimal 
SUSY breaking scenarios, where it is typically
assumed that sfermions are mass degenerate,
as required  to respect FCNC constraints \cite{susyfcnc}.
 However, mass non-degeneracy for sfermions within the same family 
but different isospin can be acceptable, 
since it will only be midly constrained by the parameter $T$. 
In fact, some  non-degeneracy can arises even in 
the minimal SUSY-GUT, by imposing universal boundary  conditions  
at the Planck scale, rather than the GUT scale \cite{polonskypom}.

 In this letter we consider
the decay $h \to gg$ (which determines the gluon fusion
mechanism) to probe the structure of soft-breaking masses predicted by 
models of SUSY breaking.
The expression for the decay width of $h\to gg$ that includes
a non-universal mass spectrum for squarks, 
can be written  as,
\begin{equation}
\Gamma(h\to gg)= \frac{G_F \alpha^2_s m^3_h }{ 4\sqrt{2} \pi }|G_{h}|^2
\end{equation}
where
\begin{equation}
G_{h^0}= G_{t,{\tilde t}} + G_{b,{\tilde b}} + G_{LR}
          + G_{UD}
\end{equation}
For $h=h^0$, $G_h$ is given by:
\begin{equation}
G_{t,{\tilde t}} = - \frac{m^2_t}{m^2_h} \frac{\cos\alpha}{\sin\beta} 
 [ f_1(\lambda_t)+ f_3(\lambda_{\tilde t_L})+ f_3(\lambda_{\tilde t_R})] 
\end{equation}
\begin{equation}
G_{b,{\tilde b} }= \frac{m^2_b}{m^2_h} \frac{\sin\alpha}{\cos\beta} 
 [ f_1(\lambda_b)+ f_3(\lambda_{\tilde b_L})+ f_3(\lambda_{\tilde b_R})] 
\end{equation}
\begin{eqnarray}
G_{LR}&=&  \frac{m^2_W s^2_w}{m^2_h c^2_W} \sin(\beta+\alpha) 
                   [Q_u (f_3(\lambda_{\tilde u_R})- 
                  f_3(\lambda_{\tilde u_L}) \nonumber \\
            & &+   Q_d (f_3(\lambda_{\tilde d_R})- f_3(\lambda_{\tilde d_L}) ] 
\end{eqnarray}
\begin{equation}
 G_{UD} =  \frac{m^2_W }{2 m^2_h c^2_W} \sin(\beta+\alpha) 
\Sigma_{u,d} [f_3(\lambda_{\tilde u_L}) - f_3(\lambda_{\tilde d_L}) ] 
\end{equation}
The expression for $h=H^0$ are obtained by making the
replacements $\cos\alpha \to -\sin\alpha$, $\sin\alpha \to -\cos\alpha$, 
$\sin(\alpha+\beta) \to -\cos(\alpha+\beta)$;
the explicit form of $f_1(z),f_3(z)$ can be found in ref.
\cite{hspira}.
The contribution proportional to the 
fermion masses are kept only for the stop and sbottom 
($G_{t,{\tilde t} }, G_{b,{\tilde b} } $),
whereas the remmaining SUSY-breaking effects are kept for all sfermions
($G_{LR}, G_{UD}$). To reduce the number
of parameters, we choose our $\mu, A_t$ values in such a way that one can
neglect L-R mixing, which was found previously to
give small effects\cite{adjouadi}.
We have used the previous equation to evaluate the contribution of 
squarks to the parameter $\epsilon_g$, for
 $m_A=200$ GeV, $\tan\beta=5,20$, and
squark masses covering the range from 200 to 700 GeV.
For simplicity,
we apply the 1-loop leading log formula to the Higgs masses and parameters.
Results are shown in table 2, and for comparision
we have also included the values obtained with universal squark masses.
These results show that the effect of non-universal masses  
can modify the $B.R.(h\to gg)$, by values of order
$\pm 20 \% $.

{\bf {6.- Conclusions.}}
We have studied the Higgs interaction with fermions and gauge bosons, 
and found that deviations from the SM prediction, 
induced by heavy quanta, can be described by a set of parameters 
$\epsilon_{f,V}$, which can be tested at future colliders.
The Higgs signal reported at LEP can exclude
a heavy fourth family, wheres Tevatron can also be used to
probe heavy quanta through the Higgs
production by gluon fusion, followed by the promising
Higgs decays into $ WW^*$.
The resulting bounds can imply the exclusion of heavy 
particles that receive their mass directly from the SM Higgs, 
including  a 4th standard or mirror family or chiral colored 
sextets and octets. 
For the 4th family case we disply a relations between
$\epsilon_{f,V}$ (the parameters that describe the corrections to
the Higgs coupling) and the Peskin-Takeuchi parameter $T$.
Within the MSSM, we also find that gluon fusion is a sensitive
probe for a non-universal spectrum of 
squarks masses, 
which allows for the possibility to test 
the sfermion masses predicted in
 the models of SUSY breaking.

\smallskip

{\small 
We thank M. Chanowitz and C. Kolda for useful discussions.
Work supported by CONACYT-SNI (Mexico). }

\bigskip

{\bf FIGURE CAPTION}

\bigskip

Fig 1. 95 \% exclusion contours for the parameter $\epsilon_g$
that can be obtained from Higgs search at Tevatron RUN-II for
Scenario-II,
with  2 (dashes) and 10 (solid) $fb^{-1}$ of integrated luminosity.
The straight (dashed) lines corresponds to $\epsilon_g=2$.

\bigskip

\bigskip

{\bf TABLE CAPTION}

\bigskip

Table 1. Values of $R_{Zb}$ obtained for
a 4th generation of quarks, for a Higgs mass
$m_h=115$ GeV (Scenario-I).

\bigskip

Table 2. Upper limits to 4th generation fermion masses
that can be obtained from Higgs search at Tevatron RUN-II,
with  10 $fb^{-1}$ (Scenario-I).

\bigskip

Table 3. Values predicted for the parameter $\epsilon_g$
for the MSSM with $\tan\beta = 5,20$. $M_{\tilde Q}$
represents the range of values taken for
$M_{\tilde U_R}$, $M_{\tilde D_L}$,  $M_{\tilde D_R}$. 

\bigskip

\bigskip

{\bf \, \, \, Table. 1} 

\begin{center}
\begin{tabular}{||l| l||}
\hline
$m_{4th}$ [GeV]  & $R_{Zb}$ \\
\hline 
\,\, 150. & \,\, 0.74 \\
\hline
\,\, 200. & \,\, 0.72 \\
\hline
\,\, 250. & \,\, 0.70 \\
\hline
\,\, 300. & \,\, 0.67 \\
\hline
\,\,  350. & \,\, 0.64 \\
\hline
\,\, 400. & \,\, 0.60 \\
\hline
\end{tabular}
\end{center}

\bigskip

\bigskip
\bigskip

{\bf \, \, \, Table. 2} 

\begin{center}
\begin{tabular}{||l| l||}
\hline
$m_h$ [GeV]  & $m^{max}_{4th}$ [GeV]\\
\hline 
\,\, 120. & \,\, 350. \\
\hline
\,\, 130. & \,\, 520. \\
\hline
\,\, 140. & \,\, 650. \\
\hline
\,\, 150. & \,\, 730. \\
\hline
\,\,  160. & \,\, 870. \\
\hline
\,\, 170. & \,\, 710. \\
\hline
\,\, 180. &\,\, 670. \\
\hline
\,\, 190. & \,\, 600. \\
\hline
\,\,  200. & \,\, 510. \\
\hline
\end{tabular}
\end{center}

\bigskip

\bigskip

{\bf \, \, \, Table. 3} 

\begin{center}
\begin{tabular}{|| l | l | l | l ||}
\hline
$m_{\tilde U}$ [GeV]  & $m_{\tilde Q}$ [GeV]  & $\epsilon_g$ (non-univ) & 
$\epsilon_g$ (univ) \\
\hline 
 200. & $200\to 700$ & \, $0.1 \to 0.4$  &  0.4  \\
\hline
 500. &  $200\to 700$ & $-0.05 \to 0.14$  &  0.05  \\
\hline
 700. &  $200\to 700$ & $-0.05 \to 0.18$  &  0.1  \\
\hline
\end{tabular}
\end{center}

\bigskip


\newpage

\begin{thebibliography}{9}  
\bibitem{higgsrev} For a recent review see: M. Kado, talk at 
Recontres de Moriond (March, 2000), hep-ex/0005022.
\bibitem{Meandthem} J.L Diaz-Cruz et al., Mod. Phys. Lett. A15 (2000)
1377 [hep-ph/9905335];
C. Kolda and L. Hall, Phys. Lett. B459 (1999) 213
[hep-ph/9904236].
\bibitem{carenaetal} For an updated calculation and references
to earlier work see: M. Carena et al., hep-ph/0001002;
see also: J.R. Espinosa and Ren-Jie Zhang, Nucl. Phys. B586 (2000) 3.
\bibitem{newsofhiggs} See for instance the talk by P. Igo-Kemenes, 
``Statuts of the Higgs boson searches''
(ALEPH, DELPHI, L3, OPAL and the LEP Higgs working group), 
Nov. 3, 2000.
\bibitem{ourhix} J.L. Diaz-Cruz et al., Phys. Rev. Lett. 80 (1998) 4641;
C. Balaaz et al., Phys. Rev. D59 (1999) 055016; M. Carena, S. Mrenna and 
C. Wagner, Phys. Rev. D60 (1999) 075010;
see also: M Carena et al, ``Report of the Tevatron Higgs working group'',
hep-ph/0010338.
\bibitem{hixhunter} J. Gunion, H. Haber, G. Kane and S. Dawson,
'The Higgs Hunters Guide', Adison Wesley, Reading 1990.
\bibitem{myhlfv} J.L. Diaz-Cruz and J. Toscano, Phys. Rev. D62 (2000) 116005 ;
M. Sher, Phys. Lett. B487 (2000) 151;
T. Han and D. Marfatia, Phys. Rev. Lett. 86 (2001) 1442.
\bibitem{chanofourth} M. Chanowitz, Phys. Rev. Lett. 69 (1992) 2037;
I.F. Ginzburg, I.P. Ivanov and A. Shiller, Phys. Rev. D60 (1999) 095001;
 S. Dawson and H. Haber, Phys. Rev. D44 (1991) 53.
\bibitem{veltmanetal} M. Veltman, Nucl. Phys. B123 (1977) 89;
M. Chanowitz, M. Furman and I. Hinchliffe, Nucl. Phys. B153 (1979) 402.
\bibitem{polonskypom} N. Polonski and A. Pomarol, Phys. Rev. D51 (1995) 6532.
\bibitem{chiralferms} J. A. Minahan, P. Ramond and B.D. Wright,
Phys. Rev. D42 (1990) 1692.
\bibitem{erlerlangack} J. Erler and P. Langacker, in Review of
Particle Properties, Eur. Phys. J C3 (1998) 1.
\bibitem{siannah} A. Dobado, M.J. Herrero and S. Penaranda,
Eur. Phys. J.  C7 (1999) 313.
\bibitem{LowETH} A. Vainshtein et al. Sov. J. Nucl. Phys. 30 (1979) 711;
B. Kniehl and  M. Spira Z. Phys. C69 (1995) 77.
\bibitem{hspira} For a review see: M. Spira, Fortsch. Phys. 46 (1998) 203. 
\bibitem{peskintak} M. Peskin and T. Takeuchi, Phys. Rev. Lett. 65 (1990) 964.
\bibitem{Hanetal} T. Han, A. Turcot and R.J. Zhang, hep-ph/9812275.
\bibitem{himssm}J. Gunion and H. Haber, Nucl. Phys. B272 (1986) 1.
\bibitem{adjouadi} A. Djouadi, Phys. Lett. B435 (1998) 101.
\bibitem{susyfcnc} A. Masiero et al., Nucl. Phys. B477 (1996) 321.


\end{thebibliography}
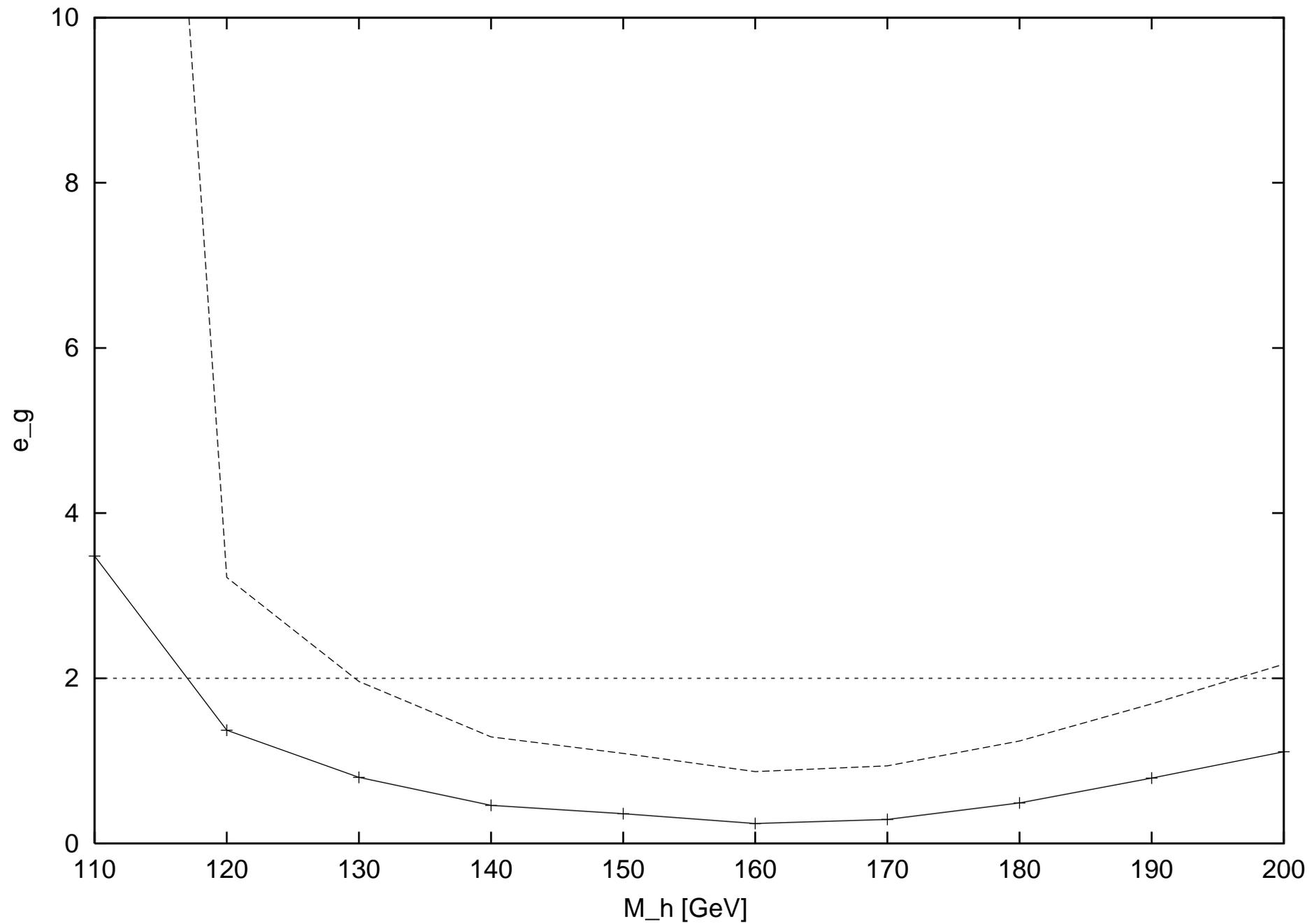
\end{document}